\documentclass[aps,pre,preprint]{revtex4}
\usepackage{graphicx}

\begin{document}
\title{Equilibrium properties of realistic random heteropolymers and their relevance for globular and naturally unfolded proteins}
\author{G. Tiana}
\affiliation{Department of Physics, Universit\`a degli Studi di Milano and INFN, via Celoria 16, 20133 Milano, Italy}
\author{L. Sutto}
\affiliation{Spanish National Cancer Research Center (CNIO), Structural Biology and Biocomputing Programme, Melchor Fernandez Almagro, 3. E-28029 Madrid, Spain}

\date{\today}
\begin{abstract}
Random heteropolymers do not display the typical equilibrium properties of globular proteins, but are the starting point to understand the physics of proteins and, in particular, to describe their non--native states. So far, they have been studied only with mean--field models in the thermodynamic limit, or with computer simulations of very small chains on lattice.  After describing a self--adjusting parallel--tempering technique to sample efficiently the low--energy states of frustrated systems without the need of tuning the system--dependent parameters of the algorithm, we apply it to random heteropolymers moving in continuous space. We show that if the mean interaction between monomers is negative, the usual description through the random energy model is nearly correct, provided that it is extended to account for non--compact conformations. If the mean interaction is positive, such a simple description breaks out and the system behaves in a way more similar to Ising spin glasses. The former case is a model for the denatured state of globular proteins, the latter of naturally--unfolded proteins, whose equilibrium properties thus result qualitatively different.
\end{abstract}


\maketitle

\section{Introduction}

Random heteropolymers are chains of molecules displaying quenched disordered interactions. Although they usually do not display a dominant equilibrium conformation as globular proteins do, they have been subject of a substantial theoretical interest. Aside from being a model for peptides built out of random sequences of amino acids, like those which are supposed to be involved in prebiotic evolution, random heteropolymers are the starting point to describe the behavior of proteins. Consequently, they have been widely used as a benchmark to study the physics of globular proteins, whose sequence is not random but has undergone natural evolution \cite{Sha:89,Bry:90,Ram:94,Pan:95}, in a way similar to that in which the ideal gas is the starting point to study real gases and the ideal chain is the starting point to study real homopolymers. 

The simplest description of a random heteropolymer is through the random energy model (REM) \cite{Der:81,Bry:87}, which assumes that the total energy of the system is the sum of a constant, large number $n_c$ of uncorrelated two--body energies, defined by an average $\epsilon_0$ and a standard deviation $\sigma$. The model predicts a parabolic entropy
\begin{equation}
S(E)=-\frac{(E-n_c\epsilon_0)^2}{2n_c\sigma^2}
\label{eq:s_e}
\end{equation}
down to a ground--state energy $E_c=zN\epsilon_0/2-N\sigma(z\log\gamma)^{1/2}$, where $z$ is the number of contacts per monomer, so that $n_c=zN/2$, and $\gamma$ is the number of conformations available to each monomer. The thermodynamics of random heteropolymers has been investigated also in the canonical ensemble with a replica approach \cite{Gut:89,Fra:92,Sfa:93}, showing the validity of the REM under the condition that only globular conformations contribute to the partition function \cite{Wil:00}.

The understanding of the physics of random heteropolymers  has been very helpful in the study of globular proteins, that is of heteropolymers which are non--random and which display at biological temperatures a unique equilibrium conformation (the "native" state). In particular, the $S(E)$ of random heteropolymers has been used to model the unfolded state of proteins. The attractive property of this $S(E)$, and of the associated $E_c$, which makes this approach useful is  its  self--averageness, namely in the limit of large $N$ it does not depend on the specific realization of the two--body energies, and consequently on the protein sequence.  If a protein sequence displays in some conformation an energy lower than the sequence--independent value of $E_c$, this is the native state of the protein and the system displays the two--state thermodynamics typical of proteins \cite{Sha:93,Gol:92}.  

This strategy has been successfully applied to lattice models,  designing proteins through the minimization of the energy of the sequence on the wanted native conformation  \cite{Sha:94}, designing potentials to fold specific sequences \cite{Sen:98,Mir:96}, estimating the effect of mutations in the protein sequence \cite{Tia:98} or studying natural evolution \cite{Tia:00}. In fact, for lattice--model heteropolymers it was shown that the free energy is indeed self--averaging \cite{Chu:01} and that the statistical independence of energy levels required by the REM, although not been strictly obeyed, is enough to allow protein design \cite{Pan:96}.

However, lattice models are not very realistic and constrain the polymer much more than what the chemistry of proteins requires. This favors the applicability of the REM, because depletes the conformational space of correlated conformations. As a matter of fact, the design of folding sequence in continuous conformational space is much more cumbersome \cite{Ama:05,Ama:06} and the attempt to obtain a potential to fold protein sequences with variational approaches of the kind of refs. \cite{Sen:98,Mir:96} has been so far frustrated, even using a strong dihedral potential to obtain a lattice--like behavior.

The goal of the present work is to investigate the thermodynamics of a heteropolymer model in continuous space, in order to understand whether the REM physics applies and what are the related consequences on the study of proteins. The model is an inextensible chain of beads put at a distance $a=3.8$ interacting through a spherical--well potential
\begin{equation}
U=\sum_{i<j-2}^N\left[B(\sigma_i,\sigma_j)\theta(R-|r_i-r_j|)+\frac{1}{\theta(R_0-|r_i-r_j|)}\right],
\end{equation}
where $B_{\alpha\beta}$ is a $20\times 20$ Gaussian random matrix with mean $\epsilon$ and standard deviation $\sigma=1$, $\sigma_i$ is a random sequence of integers in the range $[1,20]$ (to mimic natural amino acids), $\theta$ is a step function, $R=5.5\AA$ is the width of the well and $R_0=4\AA$ is the hardcore radius. 

The sampling of the conformational space of such a system is not a trivial problem, especially at the low temperatures needed to calculate $E_c$, and consequently we have first to design a computational strategy to face this problem. This is done in Section \ref{sect:sample}. In Sections \ref{sect:rsb} and \ref{sect:neg} we investigate the applicability of the REM scenario, considering that the typical size $N$ of polypetides and proteins is small (the thermodynamic limit does not make sense in this context) and that here the assumption concerning a constant number of contacts is doubtful. The coil--globule transition is discussed in Sect. \ref{sect:cg}. Particularly interesting is the case of interaction matrices with positive average, which models proteins with low content in hydrophobic amino acids \cite{Rom:01}. These proteins are intrinsically disordered and do not display a unique equilibrium state at biological temperature. In Sects.  \ref{sect:rsb} and  \ref{sect:pos} we show that they populate a phase which is physically distinct than the denatured state of proteins, and try to characterize it.

\section{The sampling algorithm: adaptive simulated tempering}
\label{sect:sample}

The study of the low--energy properties of frustrated systems is computationally challenging \cite{Parisi_book}. Replica--exchange sampling \cite{Swe:86} is a powerful technique, needs very little knowledge of the system to be sampled, but needs large parallel computers to be efficient. On the contrary, simulated tempering \cite{Mar:92} is efficient on a single processor, provided that one can tune correctly some parameters of the simulation. In fact, in simulated tempering temperature is regarded as a dynamic variable which changes in a discrete set of values $\{T_i\}$ with rate
\begin{equation}
w(T_i\to T_j)=w_0\min(1,\exp[-(1/T_j-1/T_i)E-g_j+g_i]), 
\end{equation}
where $w_0$ is the rate of attempting a te.mperature change, $E$ is the energy of the system in the current conformation and $g(T)$ are dimensionless weights which are meant to improve the diffusivity of the temperature. The system--dependent parameters to be tuned are then the set of allowed temperatures $\{T_i\}$ and the weights $\{g_i\}$. A uniform sampling of all temperatures is obtained by choosing $g_i=F(T_i)/T_i$, where $F(T_i)$ is the free energy at temperature $T_i$, which of course is not known in advance.

In order to use simulated tempering in an automatic way, we have developed an adaptive scheme which updates the values of $\{T_i\}$ and of $\{g_i\}$ in a self--consistent way, in the spirit of the approach developed in ref. \cite{Fer:02}. The idea is to carry out a simulated tempering starting at high tempertures, to estimate the density of states of the visited range of energies, and from this to obtain a lower temperature and its weight to continue the tempering efficiently. The procedure is iterated until the system reaches the desired low temperature. Moreover, at each iteration the current set of temperatures and weights is adjusted, exploiting the better knowledge of the density of states obtained as the simulation proceeds.

Specifically, the algorithm works as follows:

1) A plain Monte Carlo sampling is performed at high temperature $T_1$ for $n_{adj}$ steps, collecting the histogram of sampled energies.

2) A multiple--histogram algorithm \cite{Fer:89} extracts the density of states $g(E)$ from the histogram(s) of sampled energy. From $g(E)$, the partition functions  and the  free energies $F(T)$ are obtained by summation over the sampled energies, and the Boltzmann probability $p_T(E)$ simply from definition. In the iteration following the first one, the histograms used in the multiple--histogram are those belonging up to the 6 previous iterations.

3) A new temperature $T_i$ is added below the other(s). The new temperature is chosen in such a way that the mean--field temperature--jump rate $w_{MF}(T_{i-1}\to T_i)$ is equal to a preset  value $w_{new}$,  where
\begin{eqnarray}
&& w_{MF}(T_{i-1}\to T_i) \equiv w_0 \nonumber\\ 
&&\int dE\;\min(1,\exp[(1/T_{i-1}-1/T_i)E-g_{i-1}+g_i])\,p_{T_{i-1}}(E)
\end{eqnarray}
and the weights $g_i$ are set to $F(T_i)/T_i$.

4) If the number of temperatures $n_T$ used in the simulation is larger than 3,  the temperatures from $T_2$ to $T_{n_T-2}$ are readjusted, also allowing a decrease of $n_T$. Operatively, for each possible $n'_T>3$  the product 
\begin{equation}
w_{glob}\equiv\prod_{i=1}^{n'_T-2} w_{MF}(T_{i}\to T_{i+1}) w_{MF}(T_{i+1}\to T_i)
\end{equation}
is maximised with respect to the set $[T_2,T_{n'_T-2}]$. The minimum value of $n'_T$ such that $w_{glob}$ is larger than the value it had before is taken as new $n_T$, together with the associated temperatures and weights. 

5) A simulated tempering with the $n_T$ new temperatures and weights is carried out for a total of $n_T\cdot n_{adj}$ steps.

6) The simulated tempering is considered successful if the fraction of time that the system has spent at each temperature is larger than a preset threshold $h_t/n_T$ and if the jump probability between each pair of consecutive temperatures calculated from the simulation is larger than a preset threshold $p_t/n_T$. If the simulated tempering fails, the set of temperatures $T_i$ is substituted by the last successful set of temperatures, and the lowest temperature $T_{N_T}$ is raised to half--way with respect to $T_{N_T-1}$.

7) If the simulated tempering is considered successful, return to 2).

\begin{figure}
\includegraphics[width=8cm]{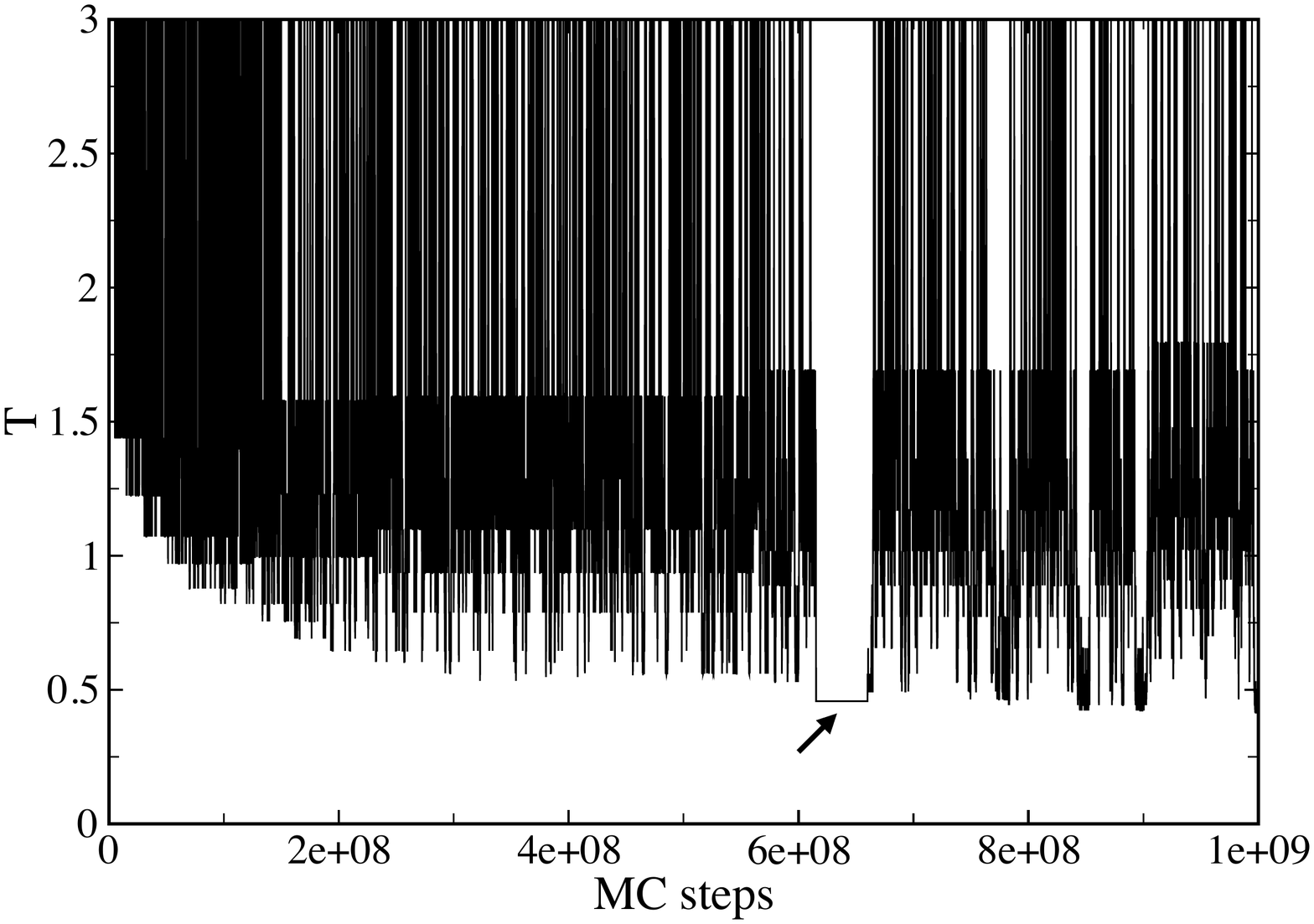}
\caption{An example of adaptive simulated tempering. The algorithm adds new temperatures as the simulation goes on. When a set of temperatures fails (see arrow), the system restarts from the last successful set, increasing the lowest temperature.}
\label{fig:stempering}
\end{figure}

Unlike standard simulated tempering, the present algorithm does not rely on the knowledge of thermodynamic features of the specific system, except for the initial temperature $T_1$. One has only to define some parameters which control the quality of the simulation, and consequently are weakly dependent on the specific system. The choice we made is $w_{new}=4\cdot 10^{-3}$, $h_t=0.1$, $p_t=0.01$, $n_{adj}=10^6$, $w_0=10^{-4}$.
An example of application of the adaptive simulated tempering is given in Fig. \ref{fig:stempering}. For each polymer and each interaction matrix we run the algorithm to obtain an efficient set of temperature and weights. Then, we repeat thrice a simulated tempering sampling keeping the temperatures fixed, to evaluate the convergence of the thermodynamic quantities. 

For each average $\epsilon$ of the interaction matrix (-1, -0.5, 0 and +1, while the standard deviation is 1, setting the energy scale of the system) and for each length $N$ of the polymer (20, 25, 30 and 60), we have sampled 20 realizations of the matrix.  Almost in all cases we could reach temperatures lower than 0.1. A typical run of the adaptive algorithm for a polymer of 60 residues takes of the order of 10 hours on a single desktop cpu. For chain of length $N=90$ we were not able to reach full equilibrium at temperatures lower than $0.7$, and consequently we discarded such simulations from the analysis.

\begin{figure}
\includegraphics[width=8cm]{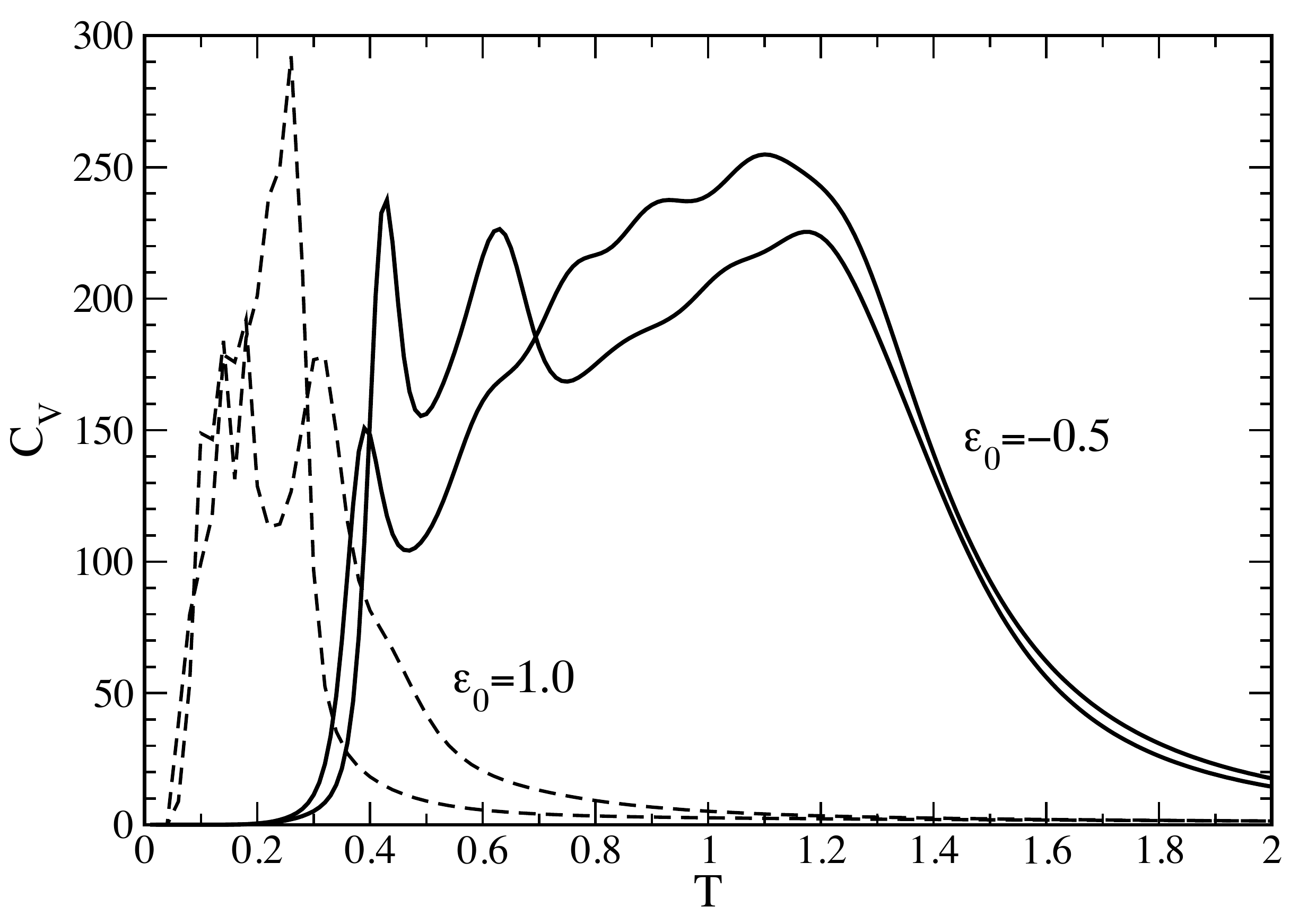}
\caption{The specific heat of a polymer interacting with two realizations of the interaction matrix at $\epsilon_0=-0.5$ and two at $\epsilon_0=+1$.}
\label{fig:cv}
\end{figure}

From these simulations one can obtain the density of states and thus all the other thermodynamic quantities (e.g. the specific heat $C_v$, as shown in Fig. \ref{fig:cv}). All $C_v$ obtained for the different realizations of the interaction matrix are quite irregular. All of them show a crowded set of peaks at very low temperaturescorresponding to the freezing of the system into the lowest available conformations. These temperatures are  $T_g\approx 0.2$ for $\epsilon_0=1$, $T_g\approx 0.4$ for $\epsilon_0=-0.5$ and $T_g\approx 0.7$ for $\epsilon_0=-1$. In all cases the simulated tempering could visit temperatures below $T_g$, as expected from the analysis of this algorithm in the case of other frustrated systems \cite{Mar:98}

\section{Interaction matrices with different average result in two different thermodynamic behaviours}
\label{sect:rsb}

The natural order parameter to study the conformational space of random heteropolymers \cite{Gut:89,Sfa:93,Pan:95} is the structural overlap
\begin{equation}
q(\alpha,\beta)=\frac{1}{\max(n_c(\alpha),n_c(\beta))}\sum_{i<j} \Delta(|r_i^\alpha-r_j^\alpha|) \Delta(|r_i^\beta-r_j^\beta|),
\end{equation}
where $n_c(\alpha)=\sum_{i<j} \Delta(|r_i^\alpha-r_j^\alpha|)$ is the number of contacts of conformation $\alpha$.

\begin{figure}
\includegraphics[width=8cm]{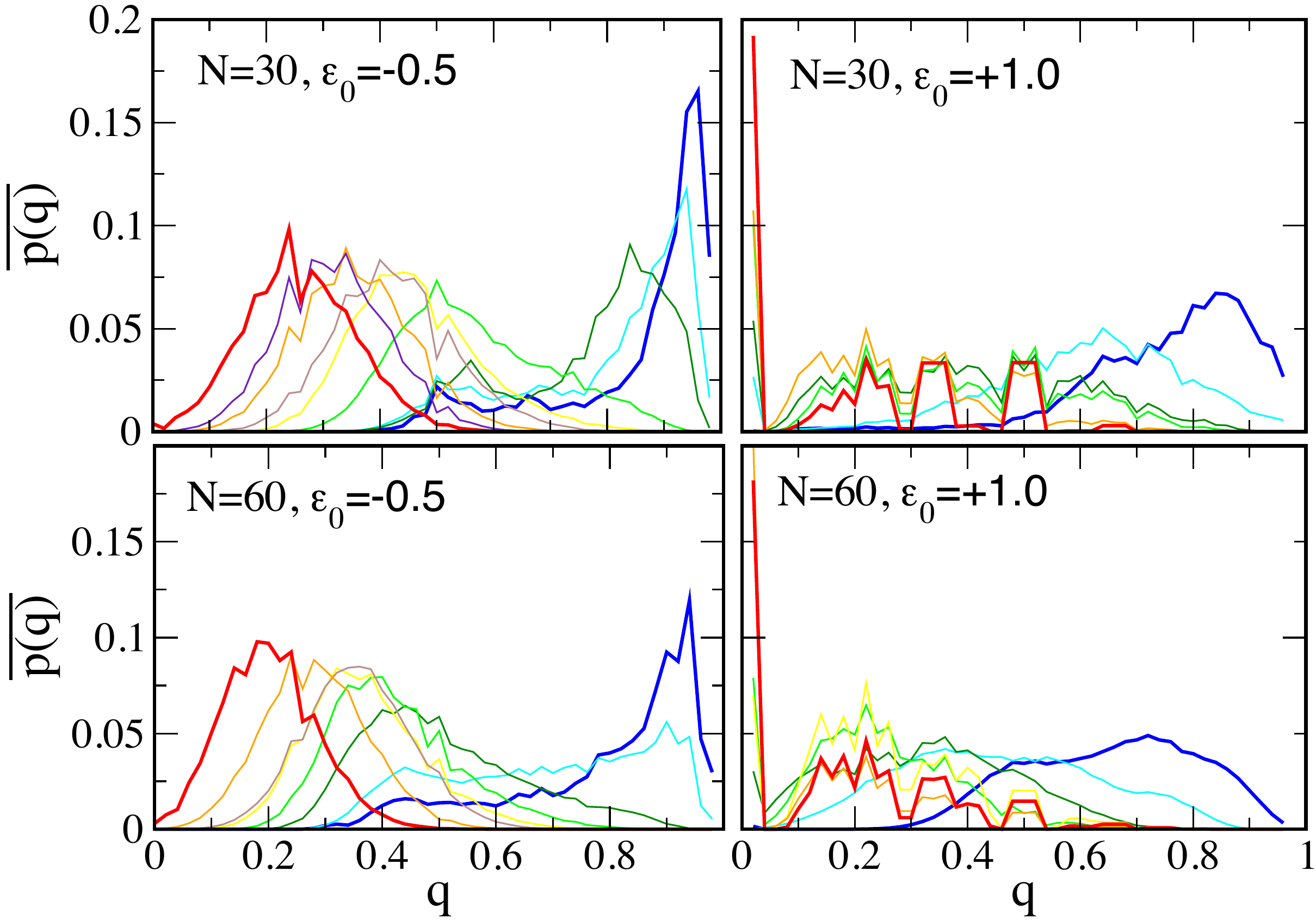}
\caption{The distribution $\overline{p(q)}$ of the order parameter for two sizes of the polymer ($N=30$ above and $N=60$ below) and for two different values of the average $\epsilon_0$ of the interaction matrix ($\epsilon_0=-0.5$ to the right and $\epsilon_0=1$ to the left). In each plot the different distributions are calculated from temperature $T=0.1$ (blue curve) to $T=1.0$ (red curve).}
\label{fig:p_q}
\end{figure}

The equilibrium distributions $\overline{p(q)}$ averaged over the interaction matrices  are displayed in Fig. \ref{fig:p_q} for some selected simulations. For negative values of the average $\epsilon$ of the interaction matrix (cf. left panels of the figure)) the distribution $p(q)$ display two well-defined peaks, one close to $q=1$ and the other one below $q=0.5$. As the temperature is decreased, the peak close to $q=1$ decreases, while the low--q peak increases and moves towards $q\approx 0.2$.

The behaviour of the distributions associated with $\epsilon_0=+1$ is qualitately different. There is a broad peak at high q, which further broadens increasing $N$. The top of the peak moves from $q=0.82$ to $q=0.74$ as $N$ increases from 30 to 60. The lower part of the distribution does not display a single peak, but a complicated pattern covering the whole range of variability of $q$. Moreover, there is a sharp peak exactly at $q=0$. The curves at $\epsilon_0=-1$ are similar to those at $\epsilon_0=-0.5$, while those at $\epsilon_0=0$ are somewhat in between the two behaviours (data not shown).

The behaviour of the distribution at $\epsilon_0=-0.5$ is typical of frustrated systems undergoing a one--step replica symmetry breaking (RSB) \cite{Parisi_book}, where thermodynamically--relevant conformations are either identical to each other (resulting in the $q=1$ peak) or markedly different (resulting in the peak at $q_0\ll 1$, cf. ref. \cite{book:FE}). On the contrary, the bulky shape of $\overline{p(q)}$ and the position of the high--q peak at $q<1$ observed in the case $\epsilon_0=+1$ suggest a more complicated pattern of RSB, more similar to the full RSB of Ising spin glasses \cite{Parisi_book,book:FE} than the one--step RSB observed in our simulations at negative $\epsilon_0$. The sharp peak at $q=0$ is a polymeric effect: increasing the temperature, the system stabilizes loose conformations characterized by few contacts (see Section \ref{sec:cg}); the probability that a pair of such conformations share their few contacts is combinatorially low. This effect is of course absent in spin glasses, where the number of contacts is fixed.

\begin{figure}
\includegraphics[width=8cm]{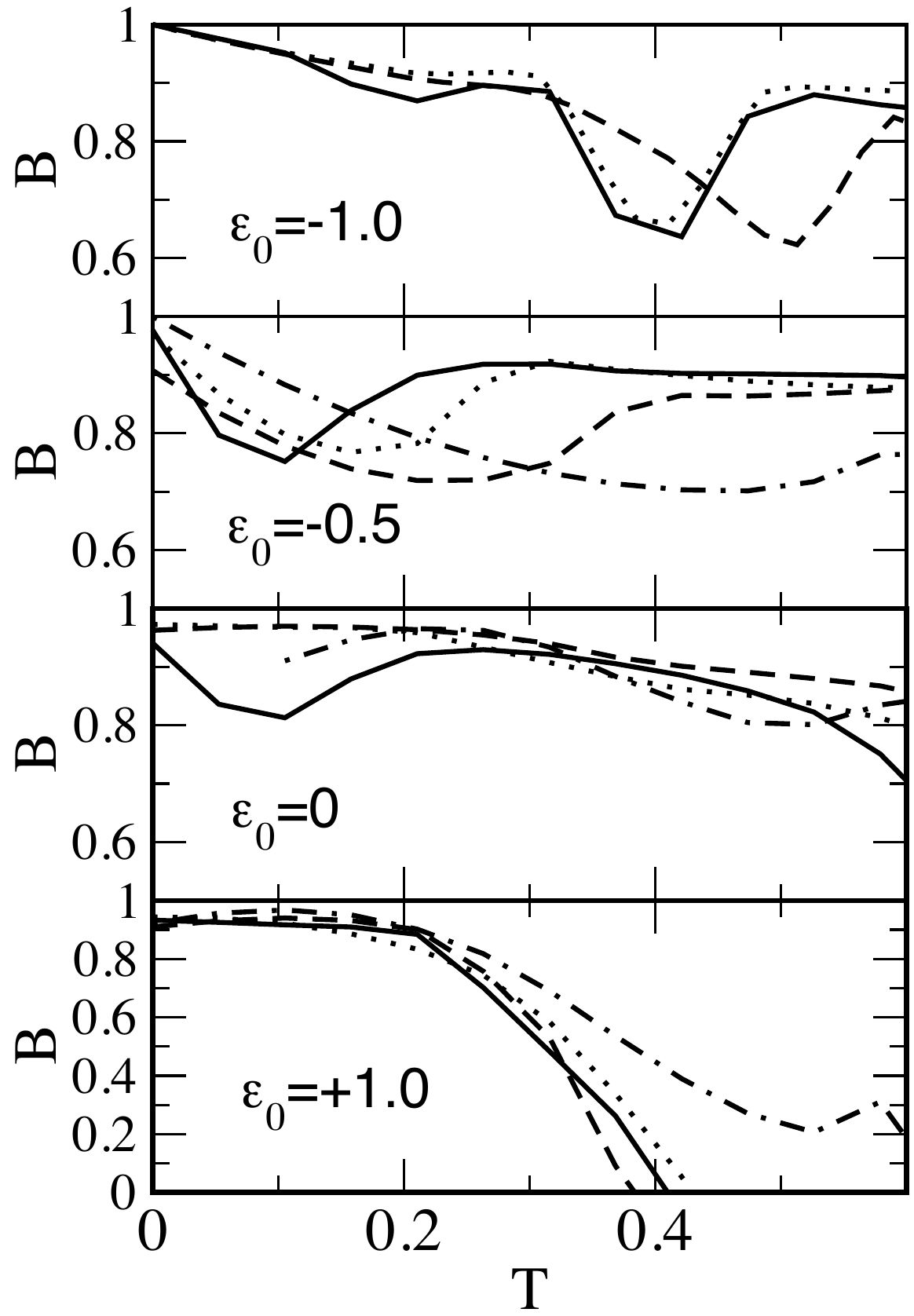}
\caption{The Binder parameter $B$, which is a measure of the kurtosis of the distribution $\overline{p(q)}$, for the polymers controlled by interaction matrices with different averages $\epsilon_0$, as a function of temperature. The curves have been smoothened to facilitate the comparison. The length of the polymer is $N=20$ (solid curve), $N=25$ (dotted curve), $N=30$ (dashed curve) and $N=60$ (dot-dashed curve). }
\label{fig:binder}
\end{figure}

A further characterization of the $\overline{p(q)}$ is given by the Binder parameter $B$, which quantifies the kurtosis of the distributions (see Fig. \ref{fig:binder}). Although one cannot draw absolute conclusions on the kind of RSB from the shape of $B(T)$ \cite{Huk:00}, it is still possible to notice that  the low-temperature part of $B(T)$ at $\epsilon_0=1$ resembles the monotonic shape of the Sherrington--Kirkpatrick model undergoing full RSB (while at higher temperatures departs from that, due to the increasing polymer--swelling effect). On the contrary, at $\epsilon_0<0$ it displays bumps similar to those of the 3--spin glass undergoing a one--step RSB \cite{Pic:01}.

\section{The $\epsilon<0$ case and its relevance for globular proteins}
\label{sect:neg}

The determination of the lowest conformational energy $E_c$ of a polymer controlled by a given interaction matrix is important to understand the folding of globular proteins. In fact, the assumption of a self--averaging behaviour of $E_c$ allow to interpret protein evolution as a minimization of their energy $E_N$ in the native conformation and describe the denatured state of the protein with the random energy model \cite{Sha:93,Sha:93b,Bry:87}.

\begin{figure}
\includegraphics[width=8cm]{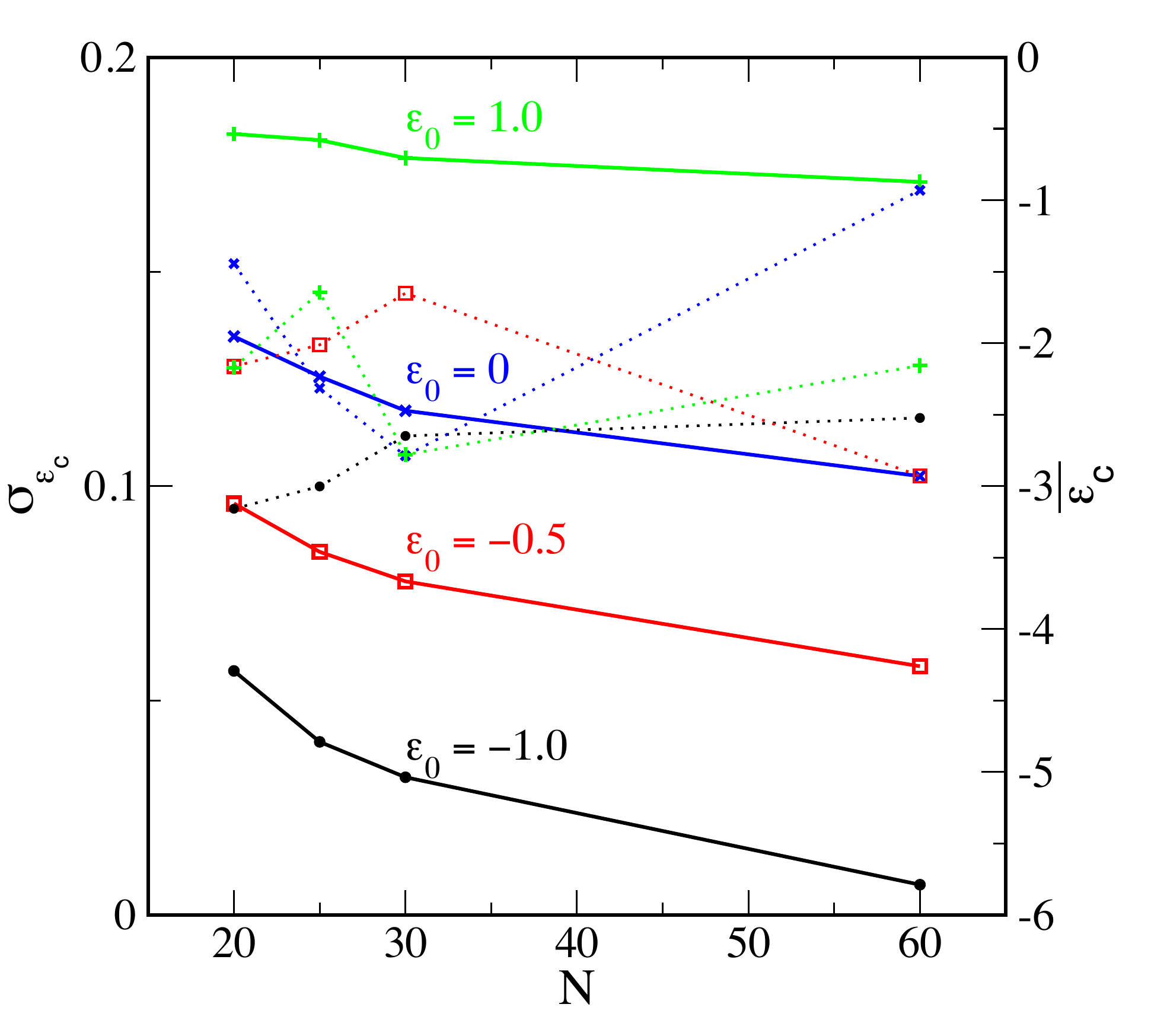}
\caption{The value of the energy density $\epsilon_c\equiv E_c/N$ of the lowest energy conformation, averaged over the realizations of the interaction matrices at different $\epsilon_0$ (solid curves, referred to the y--axis on the right), and their standard deviation (dotted curves, referred to the y--axis on the left).}
\label{fig:Ec-sigma}
\end{figure}

To verify these hypoteses, we show in in Fig. \ref{fig:Ec-sigma} the average and the standard deviation of $\epsilon_c\equiv E_c/N$ over the realization of the interaction matrices with varying values of $\epsilon_0$. The average $\overline{\epsilon_c}$ is not independent on $N$ in the range $20<N<60$, as  required by the random energy model. The curvature of the curves allow to extrapolate that  $\overline{\epsilon_c}$ becomes constant only for $N>100$. The standard deviation of $\epsilon_c$ is a few percent of the average and shows no sign of decreasing with $N$. This suggests that there is a non--negligible variability  of $E_c$ with respect to the interaction matrix and that this variability is not a finite--size effect. For instance, the typical energy scale of interaction between amino acids is $kT_{room}$ and the typical stability of globular proteins (that depends on $E_c-E_N$) is of the order of $0.1\,kT_{room}$ per residue. Consequently, the variability of $E_c$ with respect to the interaction matrix is of the same order of magnitude than the stabilization energy of the protein. Evolution has to pay an extra work to design proteins with an $E_N$ low enough to be robust against mutations and environmental changes which could affect the interaction energy. Moreover, becoming larger is not a good evolutionary strategy for proteins to become more stable, as the variability in $E_c$ would also increase.

\begin{figure}
\includegraphics[width=8cm]{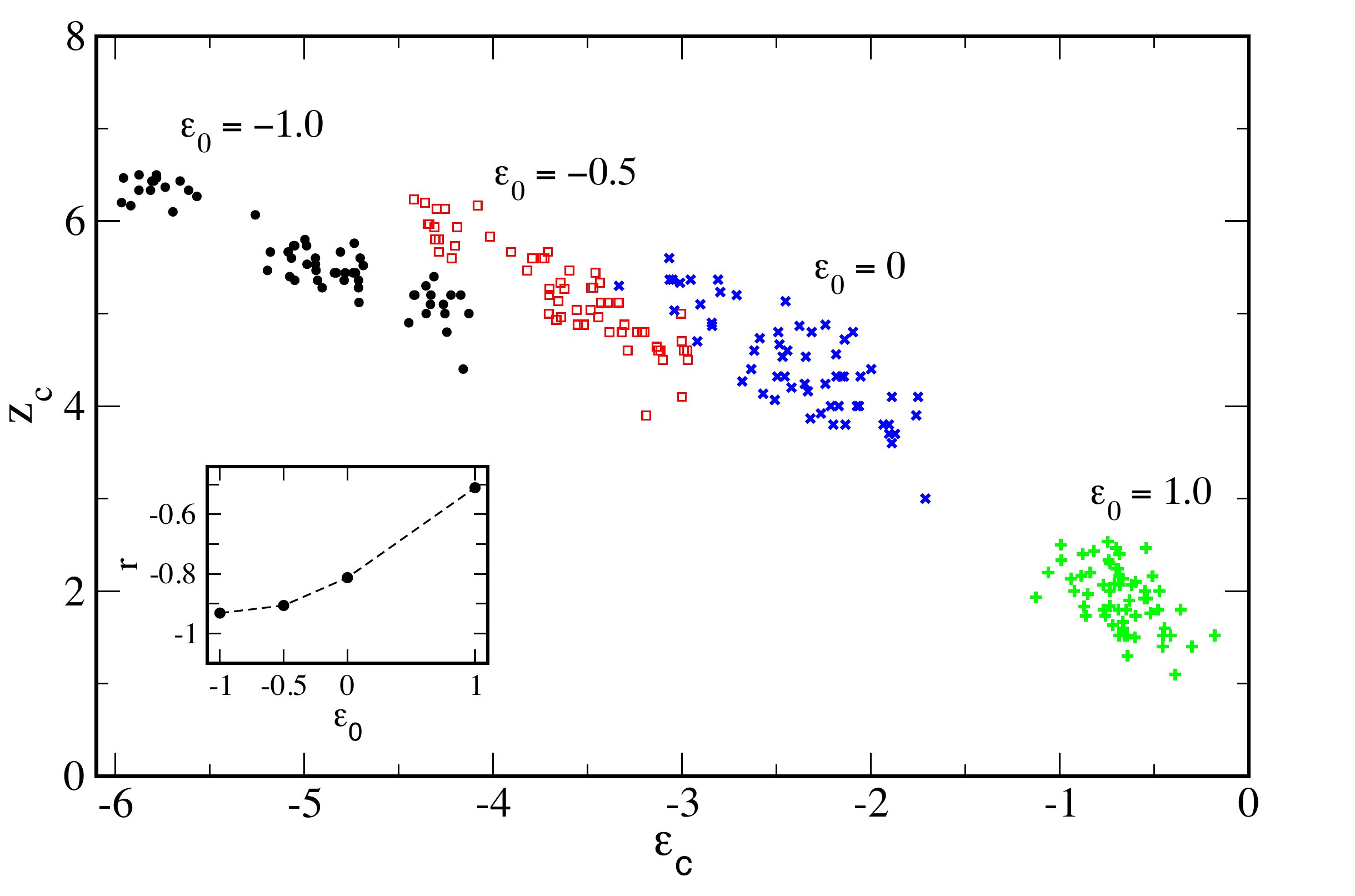}
\caption{The density of contacts $z_c$ in the lowest--energy conformation, as a function of the associated energy density $\epsilon_c$, for the different realizations of the interaction matrices at various $\epsilon_0$. In the inset, the linear correlation parameter $r$ of the data displayed in the main plot.}
\label{fig:Ec-N}
\end{figure}

To investigate the origin of the variability of $E_c$, we have plotted the density of contacts $z=2n_C/N$ within the lowest-energy conformation with respect to its energy density $\epsilon_c$ (see Fig. \ref{fig:Ec-N}). First of all, one can notice that even for $\epsilon_0<0$ the density of contacts varies within a $30\%$ from polymer to polymer, indicating that the ground state conformation is not completely compact. In this case, the value of $\epsilon_c$ is well correlated to $z$ (cf. the correlation coefficient in the inset of  \ref{fig:Ec-N}), suggesting  that the variability in $E_c$ is due to the different number of contacts that low--energy conformations can accommodate.

\begin{figure}
\includegraphics[width=8cm]{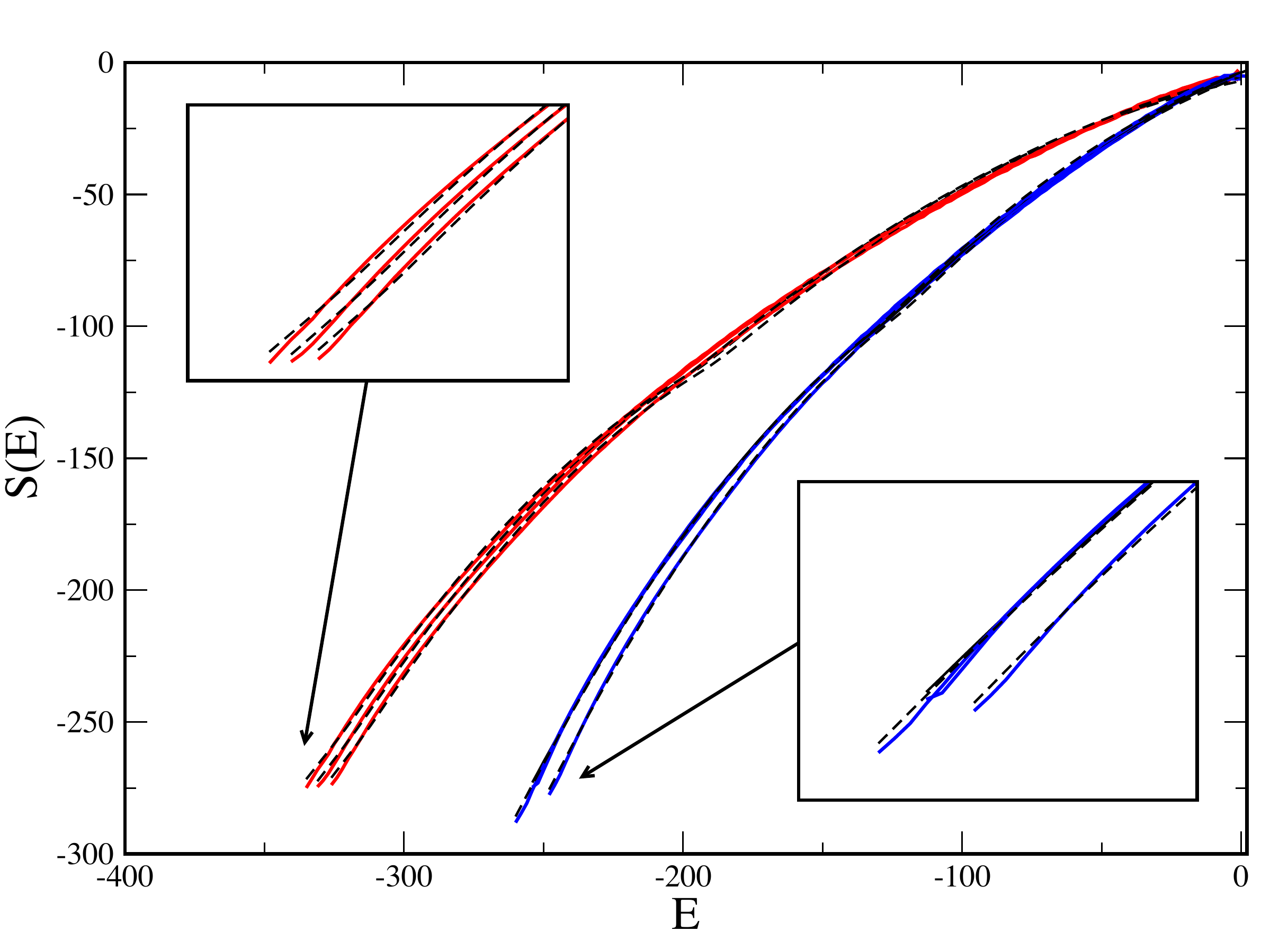}
\caption{The entropy $S(E)$ as a function of energy for three realizations of random heteropolymers of 60 monomers with average interaction energy $\epsilon_0=-1$ (red solid curves) and $\epsilon_0=-0.5$ (blue solid curves). The dashed curves indicate the fit (see text). In the inset, a zoom of the low--energy tails.}
\label{fig:s_e}
\end{figure}

The entropy function whose endpoint is $E_c$ is given in Fig. \ref{fig:s_e} for few representative cases. The curve $S(E)$ is variable with respect to the different realizations of the interaction matrix varies in the whole range of energies. Differently than the predictions of the REM, $S(E)$ cannot be fitted by the parabola of Eq. (\ref{eq:s_e}). As a matter of fact, the REM assumes a constant number of contacts, while polymers display a non-trivial distribution of the number of contacts, which is also responsible for the coil--globule transition \cite{Tay:10}.

The shape of $S(E)$ seems to display two different behaviors at low and high energies, which can be captured by a two--REM description, that is
\begin{equation}
S(E)=\log\left[\alpha\exp\left(-\frac{(E-zN\epsilon_0/2)^2}{zN\sigma^2}\right) + \alpha'\exp\left(-\frac{(E-z'N\epsilon'_0/2)^2}{z'N\sigma'^2}\right)   \right],
\label{eq:s_e2}
\end{equation}
where $N$ is the length of the chain, $\epsilon_0=-1,-0.5$ and $\sigma=1$ are the mean and the standard deviations of the interaction matrices. The fit over the other parameters gives a reduced $\chi^2$ which ranges between $0.98$ and $3.07$ for $N=60$ and worsen for smaller chains. This suggests that it is possible to describe effectively the states of the heteropolymer as a superposition of two set of conformations building out, respectively, $z$ and  $z'$ interactions per monomer, and corresponding to the two gaussians in Eq. (\ref{eq:s_e2}). The fit gives values of $z$ and $z'$ whose means over the realizations is $\overline{z}=3.22\pm 0.12$ for $\epsilon_0=-0.5$ and $\overline{z}=4.10\pm 0.14$ for $\epsilon_0=-1$. This indicates compact conformations, but not fully compact (becuase $\overline{z}$ increases with $\epsilon_0$). Moreover, the variability over the different realizations of the interaction matrix is very small. On the other hand, to the high--energy part of $S(E)$ is associated a $\overline{z'}=1.52\pm 0.35$ and $\overline{z'}=0.34\pm 0.22$ for $\epsilon_0=-0.5$ and $\epsilon_0=-1$, respectively. These correspond to more swollen conformations and suffer a much larger variability among realizations of the interaction matrix. To fit properly the high--energy part of the computed $S(E)$ it is also necessary to use values of $\epsilon'_0$ and $\sigma'$ different from the  values set for the interaction matrix (the fit with the actual values $\epsilon'_0=-1.-0.5$ and $\sigma'=1$ give a reduced $\chi^2>30$).

The picture which emerges is that for negative values of $\epsilon$ the REM fails because the underlying hypothesis of a constant number of contacts does not apply. Nonetheless, the system can be effectively described as built out of two set of conformations, different for the (constant) number of contacts, each of them displaying a REM--like behavior. The parameters controlling the more--compact conformations are rather independent on the interaction matrix, while those controlling the swollen conformations are not.

The low--energy end of the $S(E)$ curve is irregular and departs slightly from the REM behaviour in a matrix--dependent way (see insets of Fig. \ref{fig:s_e}). This irregularity involves only an interval of few $\sigma$ and sets the actual value of $E_c$ below that predicted by the REM. Thus, the major determinant in the variability of $E_c$ seems to be the shape of the high--energy part of $S(E)$.

\section{The coil--globule transition} \label{sec:cg}
\label{sect:cg}

\begin{figure}
\includegraphics[width=8cm]{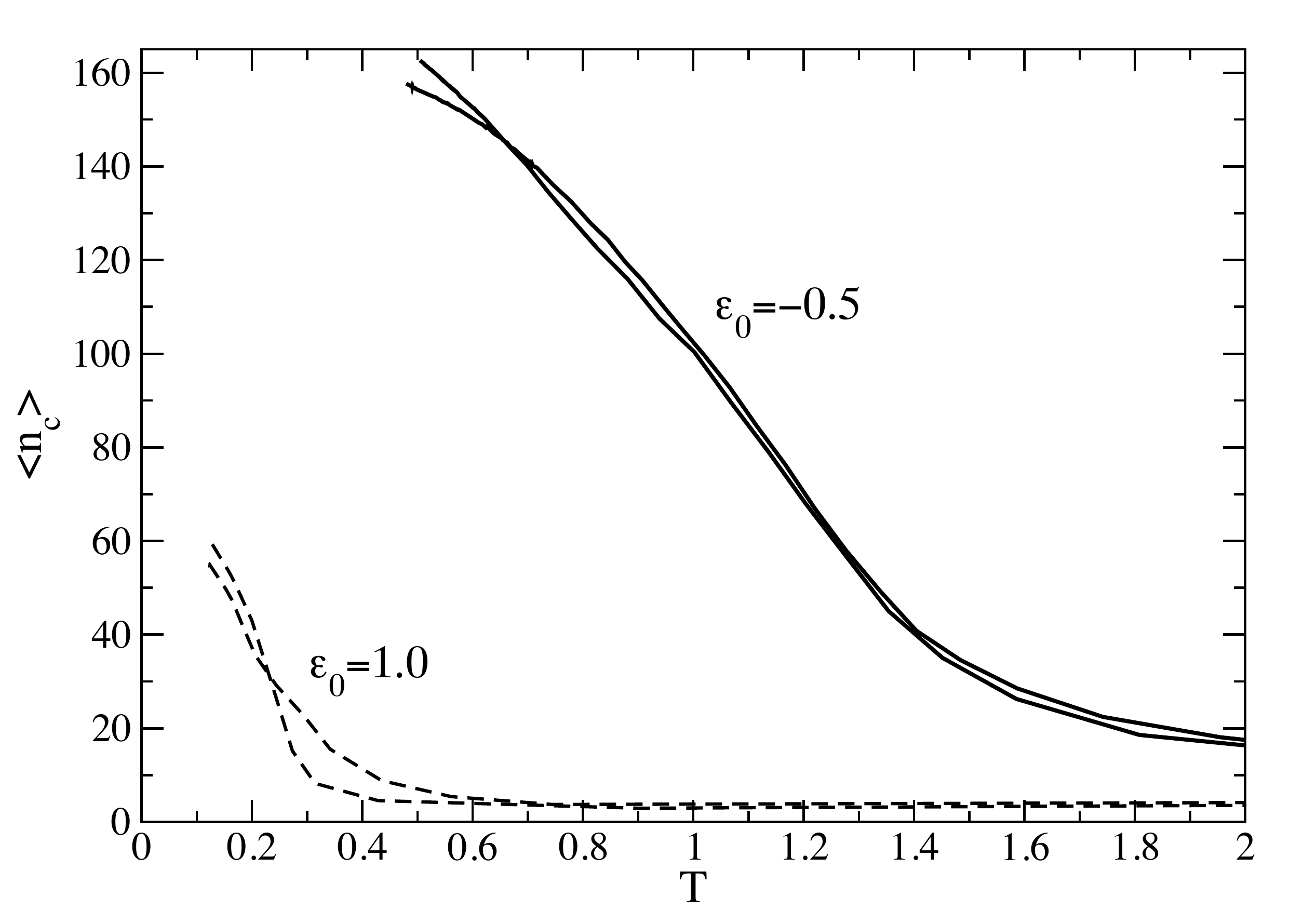}
\caption{The average number of contacts with respect to temperature.}
\label{fig:nc}
\end{figure}

The heteropolymers interacting with $\epsilon_0<0$ display a broad peak in the specific heat at high temperature (see Fig. \ref{fig:cv}) which corresponds to the midpoint in the decrease of the number of contacts in the chain (see Fig. \ref{fig:nc}), and consequently to a transition from globular to coil states. In the protein--like range of chain lengths the width of the peak in $C_v$ is quite constant with respect to the realization of the interaction matrix (e.g., at $\epsilon_0=-0.5$ $\Delta T=0.78\pm0.06$ at half height for $N=60$ and $\Delta T=0.82\pm0.14$ for $N=30$) but does not display the $N^{-1/2}$ behavior associated to the coil--globule transitions of homopolymers.

\begin{figure}
\includegraphics[width=8cm]{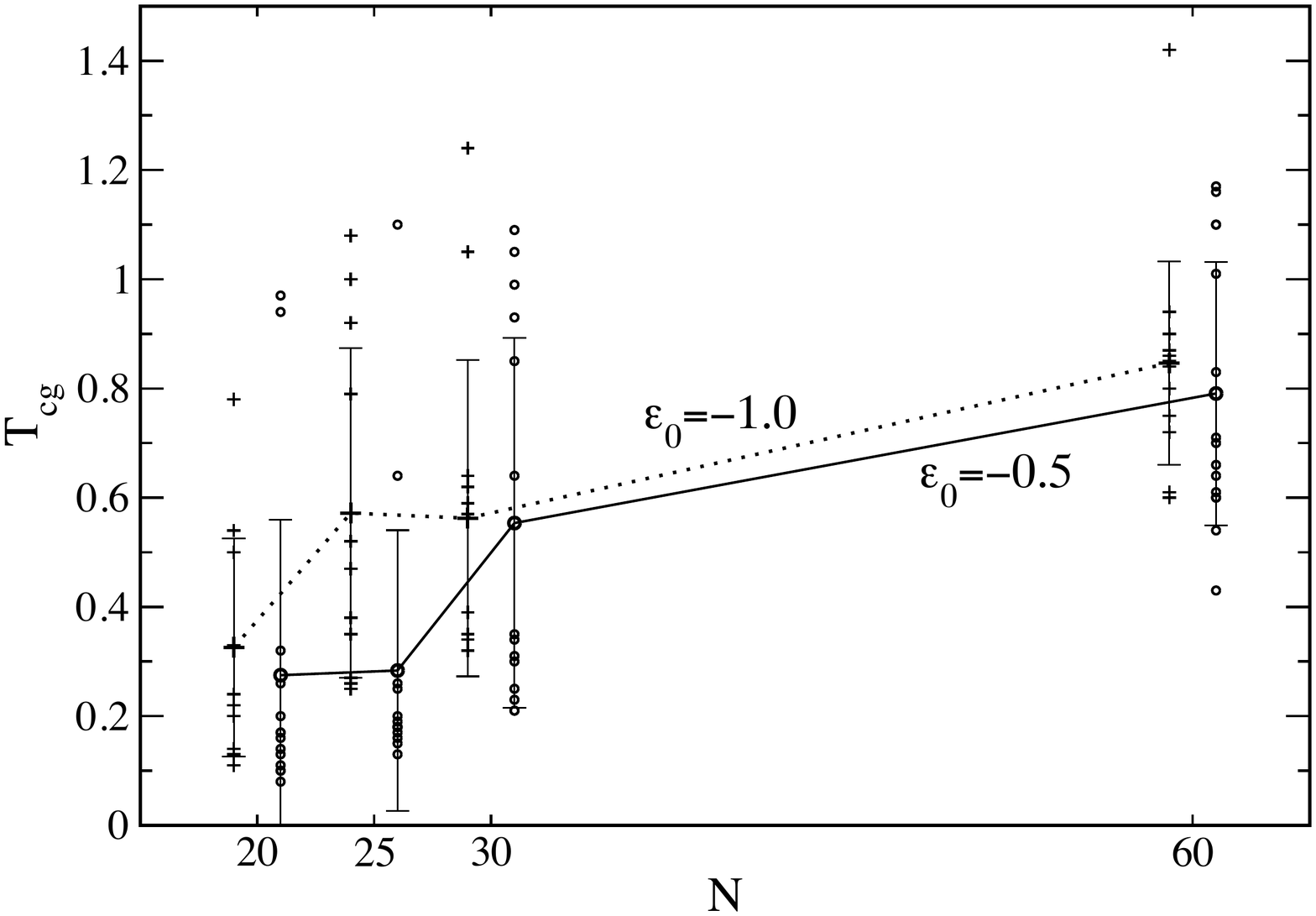}
\caption{The coil--globule transition temperatures for the different realizations of the interaction matrix with $\epsilon_0=-1$ (crosses) and $\epsilon_0=-0.5$ (circles). The dotted and solid lines indicate the average over realizations, while the error bars the associated standard deviation. The set of points at different values of $\epsilon_0$ and a given value of $N$ are slightly displaced along the horizontal axis to allow a clearer identification.}
\label{fig:tcg}
\end{figure}

On the other hand, the transition temperature (operatively defined as the temperature $T_{cg}$  corresponding to the top of the highest-T peak in $C_v$) is rather realization--dependent, and does not show any clear trend in decreasing its variability with respect to the length of the heteropolymer, as shown in Fig. \ref{fig:tcg}.

The replica approach of ref \cite{Sha:89b} highlights a coil--globule transition only for $\epsilon_0>0$, due to the strong hypotheses on the density of the chain. A Flory--Huggins description of a homopolymer corrected with the effective second virial coefficient introduced in ref. \cite{Sha:89b} predicts a coil--globule transition which depends on the number of contacts of the most compact conformation \cite{Ama:06}. The values of $T_{cg}$ calculated according to ref.  \cite{Ama:06} from the number of contacts displayed in Fig. \ref{fig:Ec-N} correlate poorly ($r=0.32$) with the values of  $T_{cg}$ of Fig \ref{fig:tcg}. This suggests that simple mean--field theories are not able to capture the large variability of the coil-globule transition temperatures.

A high--temperature--expansion approach identifies a first--order transition between a frozen globule and a random coil, and a second--order transition between a random globule and a coil \cite{Pan:97}. Due to the limited size of the heteropolymers which can be treated computationally, we are not able to investigate the order of the transition. 

A striking feature which emerges from these calculations is that the coil--globule transition is very broad in the range of polymer lengths corresponding to single--domain proteins. This means that it is very likely that the denatured state of proteins belong to the transition region, which also depends on the details of the interaction between amino acids. As a consequence, one expects a high variability in the size of the denatured state of proteins.

\section{The $\epsilon>0$ case and its relevance for natively unfolded proteins}
\label{sect:pos}

\begin{figure}
\includegraphics[width=8cm]{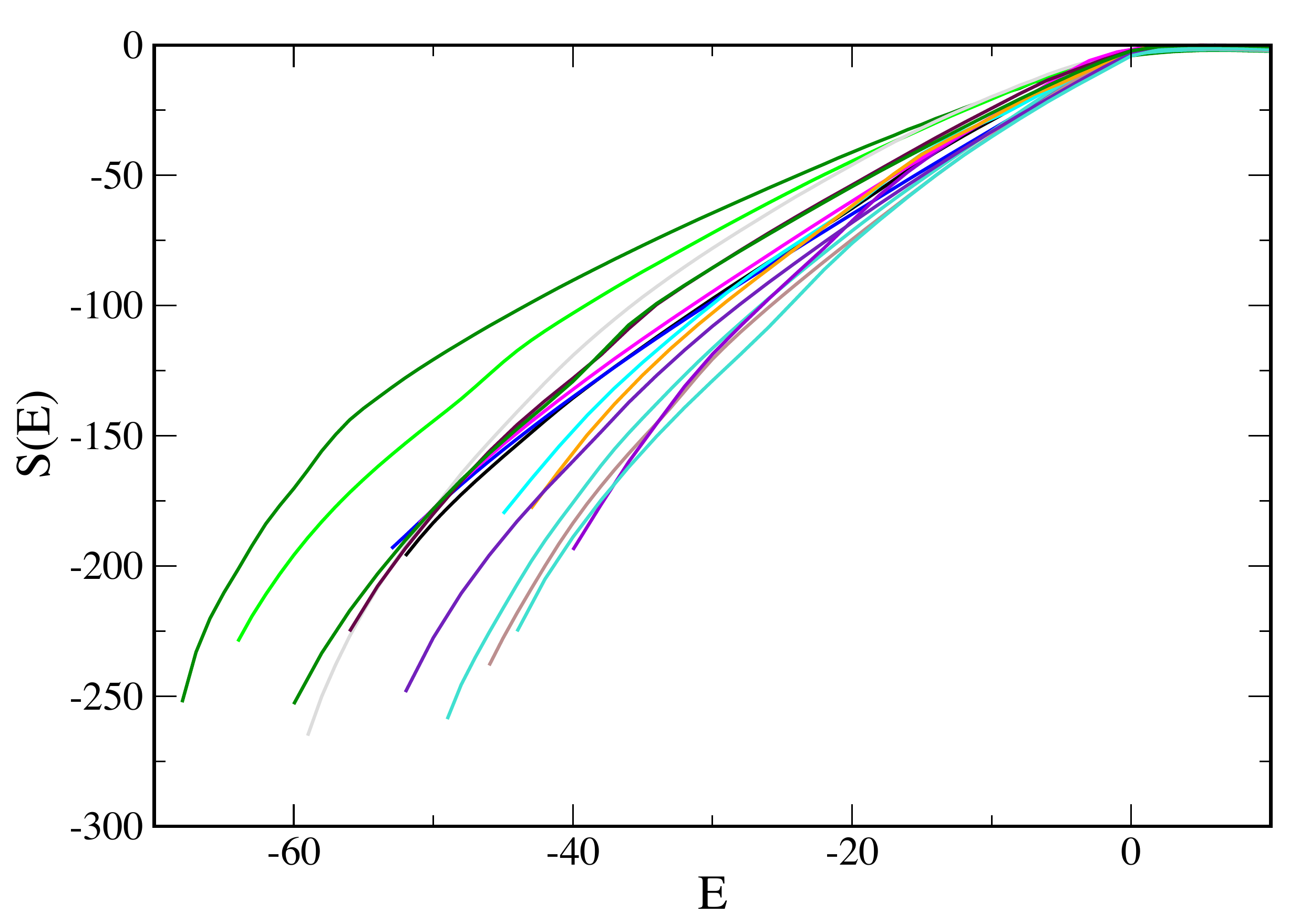}
\caption{The entropy $S(E)$ as a function of energy for fifteen realizations of random heteropolymers of 60 monomers with average interaction energy $\epsilon_0=1$.}
\label{fig:s_e1}
\end{figure}

As discussed in Sect. \ref{sect:rsb}, in the case of positive mean of the interaction matrix the REM scenario does not hold. In fact, the shape of $S(E)$ associated with the different realizations of the matrix with mean $\epsilon_0=1$ displays a irregular behavior (see Fig. \ref{fig:s_e1}). These curves cannot be fitted by a parabola or by Eq. (\ref{eq:s_e2}), the reduced $\chi^2$ being larger than 30, in agreement with a complicated RSB pattern.

The ground--state conformations display a spread which is comparable with that of negative $\epsilon_0$ (cf. Fig. \ref{fig:Ec-sigma}), something which is somewhat unexpected, due to the irregularity of the associated $S(E)$. Such conformations are still globular (with a $z$ of the order of 2, see Fig. \ref{fig:Ec-N}) even if much less compact than those with $\epsilon_0<0$. But differently from that case, now the number of contacts decreases drastically above $E_c$. From a canonical--ensemble point of view, the specific heat displays a single broadened (and quite irregular) peak at low temperatures (see Fig. \ref{fig:cv}), which inevitably marks the glassy transition. But in the same range of temperatures ($T=0.1-0.4$) the average number of contacts decreases to values typical of coils, consequently the system jumps from a glassy globule to a random coil.

The study of random heteropolymers interacting through a matrix with positive mean is interesting because they represent the lowest--order approximation of natively unfolded proteins. This class of proteins do not display a unique native conformation in solution, but display biological activity either when unstructured, or getting structured (or partially structured) upon binding other molecules \cite{Uve:02}. The low content of hydrophobic amino acids in this kind of proteins \cite{Rom:01} suggests that the average interaction is much less attractive than that of globular proteins. The hydrodynamics radius of natively unfolded proteins in solution is either that of a random coil or that of a molten globule, depending on the specific protein \cite{Uve:02}. 
This indicates that biological temperature is likely to lie in or close to the transition region ($T=0.2-0.4$ in Fig. \ref{fig:nc}), also in the case of coil proteins which get promptly compacted upon binding.

A consequence of this scenario is that the free--energy profile of natively unfolded proteins is quite different from that of unfolded globular proteins. As in Ising spin glasses, where a full RSB transition applies, the free energy displays a hierarchical tree of states at all energy scales, giving rise to conformational substates reminiscent of those observed in myoglobin but at very low temperature \cite{Fra:88}. In other words, while the denatured state of globular proteins is expected to display a set of conformations which can only be completely different from each other (i.e., are different at the length scale of the whole protein), natively unfolded proteins are expected to populate conformations which are different from each other at all possible length scales.

As a matter of fact, several experiments carried out on alpha-synuclein, a natively unfolded protein, give results which are consistent with this picture. Single--molecule F\"orster resonance energy transfer experiments at room temperature provide broadened distributions of distances between pairs of residues \cite{Tre:09}. While the authors of this work comment that {\it (they) cannot eliminate the possibility that peak broadening results from alpha--synuclein sampling of two or more specific conformations with close mean energy--transfer--efficiency values that are not resolved in our measurement}, a full RSB scenario would explain well such a peak broadening. A study of the same protein with fluorescent energy transfer combined with electron transfer measurement also shows  {\it distance distributions that are singlificantly broadened and appear to be continuous at the fitting resolution of 2 \AA} \cite{Lee:05}, compatibly with the scenario we suggest.

Moreover, nuclear magnetic resonance experiments on another natively--unfolded protein, HIV-1 Tat, show many weak crosspeaks, more abundant and more broadened than expected from a 92--residue protein \cite{Sho:06}. Again, this is compatible with an energy landscape whose roughness involves all length--scales.

The full--RSB scenario has deep consequences in the way one models the experimental data in order to obtain structural information on natively unfolded protein. In fact, one approach is to try to obtain well-defined clusters of conformations which are overall compatible with the data \cite{Mit:10}. If a one--step RSB applied, such clusters would be naturally defined as different on the length scale of the whole protein. But in a full--RSB scenario, where differences apply over a continuum of length scales, one expects the classification of probable conformation to be more complicated, due to the absence of a natural length scale to distinguish between them.

Another interesting phenomenon which takes place in frustrated systems is  the existence of slow and multiple relaxation time scales and ageing \cite{book:FE}. It was shown that these phenomena affect both RSB schemes, although at different scales \cite{Ben:08}. Anyway, they have been observed in the case of natively--unfolded proteins on the time--scale of milliseconds to microseconds \cite{Sho:06,Wu:08}, and not in the denatured state of globular proteins up to the time scale of nanoseconds \cite{Net:08}, suggesting a marked quantitative difference between the two cases.

\section{Conclusions} 

The study of random heteropolymers with continuous degrees of freedom and with size comparable to that of small globular proteins is computationally demanding, but important to complement the available mean--field theories and the old simulations done with lattice models and very short polymers.  
If the interaction energy between monomers is negative, random heteropolymers display a one--step replica symmetry breaking at low temperatures, and the density of states can be described by a modified random energy model which accounts also for non--compact conformations. The effect of these swollen conformations is to cause a variability in the lowest energy $E_c$ available to the system with respect to the details of the interaction potential.  Such a variability is not negligible and should be accounted in the {\it de--novo} design of globular proteins and in the study of the effect of mutations in protein sequences..

On the other hand, if the mean interaction energy is positive, a more complicated replica symmetry breaking pattern takes place at low temperatures, suggesting that the free--energy profile is now more complicated. This scenario is expected to apply to natively--unfolded proteins, whose residues are in average less hydrophobic than those of globular proteins.

We think that the self--adjusting simulated--tempering algorithm employed for this investigation can be useful to study the low--energy properties of other systems without the need of massively parallel computers.



\newpage

\end{document}